\documentclass[]{IEEEtran}

\usepackage[ruled]{algorithm2e}
\usepackage{stfloats}
\usepackage{bbm}
\usepackage{amssymb}
\usepackage{amsmath,amsthm,amssymb}
\usepackage{graphicx}
\usepackage{subfigure}
\usepackage{booktabs} 
\usepackage{epsfig,epsf,color,balance,cite}
\usepackage{threeparttable}
\usepackage{verbatim}
\usepackage{url}
\usepackage{booktabs}
\usepackage{bm}
\usepackage{caption}
\usepackage{diagbox}
\usepackage{multirow}
\usepackage{enumerate}
\usepackage{pifont}
\usepackage{tablefootnote}
\DeclareMathOperator*{\argmin}{arg\,min}

\begin{document}

\title{Channel-Aware Deep Learning for Superimposed Pilot Power Allocation and Receiver Design}
\author{\IEEEauthorblockN{Run Gu\textsuperscript{1,2}, Renjie~Xie\textsuperscript{3}, Wei Xu\textsuperscript{1,2}, Zhaohui~Yang\textsuperscript{4}, and Kaibin~Huang\textsuperscript{5}}

 \IEEEauthorblockA{
 \textit{\textsuperscript{1}National Mobile Communications Research Laboratory, Southeast University, Nanjing, China} \\
 \textit{\textsuperscript{2}Purple Mountain Laboratories, Nanjing, China} \\
 \textit{\textsuperscript{3}School of Internet of Things, Nanjing University of Posts and Telecommunications, Nanjing, China}\\
 \textit{\textsuperscript{4}College of Information Science and Electronic Engineering, Zhejiang University, Hangzhou, China}\\
\textit{\textsuperscript{5}Department of Electrical and Electronic Engineering, The University of Hong Kong, Hong Kong}\\
% \textit{\textsuperscript{5}Department of Electrical and Computer Engineering, The Ohio State University, USA}\\
 Emails: \{rung, wxu\}@seu.edu.cn, renjie\_xie@njupt.edu.cn, yang\_zhaohui@zju.edu.cn, huangkb@hku.hk}
  }
  
	\maketitle
	\pagestyle{empty}
	\thispagestyle{empty}
\maketitle

\begin{abstract}
Superimposed pilot (SIP) schemes face significant challenges in effectively superimposing and separating pilot and data signals, especially in multiuser mobility scenarios with rapidly varying channels. To address these challenges, we propose a novel channel-aware learning framework for SIP schemes, termed CaSIP, that jointly optimizes pilot-data power (PDP) allocation and a receiver network for pilot-data interference (PDI) elimination, by leveraging channel path gain information, a form of large-scale channel state information (CSI). The proposed framework identifies user-specific, resource element-wise PDP factors and develops a deep neural network-based SIP receiver comprising explicit channel estimation and data detection components. To properly leverage path gain data, we devise an embedding generator that projects it into embeddings, which are then fused with intermediate feature maps of the channel estimation network. Simulation results demonstrate that CaSIP efficiently outperforms traditional pilot schemes and state-of-the-art SIP schemes in terms of sum throughput and channel estimation accuracy, particularly under high-mobility and low signal-to-noise ratio (SNR) conditions.
\end{abstract}


% \fi
\IEEEpeerreviewmaketitle

\section{Introduction}	
%\IEEEPARstart {E}{hancing} spectral efficiency is a persistent objective in the advancement of wireless communication systems~\cite{xw2023toward}. Advanced techniques such as massive multiple input multiple output (MIMO) and orthogonal frequency division multiplexing (OFDM) have significantly contributed to achieving this goal. However, the effectiveness of these techniques hinges on accurate channel state information (CSI), which is typically acquired through the transmission of pilot symbols. In traditional pilot schemes, the pilot symbols occupy dedicated resources, which, while ensuring localized CSI estimation accuracy, inherently leads to a trade-off between pilot and data transmissions due to their competition for limited wireless resources in the air interface. To address this issue, superimposed pilot (SIP) techniques have emerged as a promising alternative. 

\IEEEPARstart {E}{nhancing} spectral efficiency is a persistent objective in the advancement of wireless communication systems~\cite{xw2023toward}. Superimposed pilot schemes embed pilot symbols into data symbols within the power domain, enabling their simultaneous transmission over shared resources. This approach precludes resource contention and thus improves spectral efficiency. 

However, implementing the SIP schemes raises two fundamental concerns: how to superimpose and separate pilot and data symbols. On one hand, the superimposition of pilot and data symbols can be characterized by pilot-data power (PDP) factors, which have the same dimensionality as the data symbols and define the power levels to the pilot and data symbols. The resulting superimposed symbols typically satisfy a unit-power constraint, which ensures consistent signal strength across resources for improving transmission reliability and reducing potential power imbalances. It is worth mentioning that PDP factors are pivotal in balancing channel estimation and data detection performance. Specifically, allocating higher power to pilot symbols improves channel estimation accuracy, while allocating more power to data symbols enhances data detection reliability. On the other hand, separating data symbols from the superimposed symbols requires addressing pilot-data interference (PDI) at the SIP receiver. During channel estimation, interference from data symbols can degrade performance, while during data detection, interference from pilot symbols poses challenges.

Numerous studies have explored PDP factor optimization and PDI elimination to validate the potential of SIP schemes for enhancing spectral efficiency. Studies grounded in Shannon capacity analysis have identified optimal PDP factors using various approaches, such as closed-form derivation~\cite{haritha2024superimposed}, sub-optimal iterative solvers~\cite{zhou2024optimized}, and exhaustive search approaches~\cite{xie2024superimposed}. However, these methods typically assume block-fading channel models, where the channel coherence time exceeds the symbol transmission duration. In complex scenarios with rapidly varying channels, these analytical and optimization methods may struggle to provide optimal PDP factors. For instance, in~\cite{muntan2024optimal}, the authors analyzed channel estimation errors in frequency-selective channels from 5G New Radio (NR) standards, but employed empirically determined PDP factors. Additionally, in practical SIP-assisted communication systems involving signal reception, optimal PDP factors are commonly determined through numerical simulations~\cite{jing2018superimposed,gan2024bayesian,qian2023enhancing}. Importantly, these SIP receiver designs have demonstrated effective iterative channel estimation and data detection (ICEDD) methods for PDI elimination, leveraging data estimates to enhance subsequent channel estimation. In~\cite{jing2018superimposed}, the authors alternated between a Tikhonov regularization-based channel estimation module and a matched-filter data detection module. In~\cite{gan2024bayesian}, an expectation maximization-based Bayesian learning approach was employed for the iterative framework, where digital modulations were utilized to improve data estimate accuracy. In~\cite{qian2023enhancing}, the iteration process was extended to include channel estimation, detection, and decoding modules, where error correction codes helped alleviate the adverse effects of data interference. However, these iterative SIP receiver approaches generally assume predetermined PDP factors uniformly applied across all users and resources, which decouples PDP factor optimization from PDI elimination. Furthermore, they often face high computational complexity when dealing with complex channel conditions.

Deep learning (DL) algorithms offer promising solutions to these challenges owing to their proficiency to address nonlinear signal distortions~\cite{xie2023disentangled,zeng2023multi} and facilitate end-to-end optimization~\cite{xu2023edge,xu2024disentangled}. Inspired by these strengths, DL-based SIP approaches have been proposed to jointly optimize PDP factors and SIP receivers for effective PDI elimination. For instance, in~\cite{aoudia2021end}, the resource-element (RE)-wise PDP factors and a data-driven neural receiver were jointly optimized in single-user scenarios. In multiuser MIMO systems, a convolutional layer-based network was introduced in ~\cite{gu2024learning} to optimize user-specific, RE-wise PDP factors while refining channel estimation. Nevertheless, a single iteration between channel estimation and data detection was necessary for satisfactory data detection performance.

Concurrently, environment-aware communication has emerged as a promising paradigm in next-generation wireless networks, which leverages communication environmental information, such as transceiver positions and radio channel characteristics, to facilitate CSI acquisition~\cite{zeng2024tutorial}. This capability holds particular significance for SIP schemes, where channel estimation is severely hindered by PDI. 
%Advancements in wireless networks have made it possible to acquire high-quality channel information through techniques such as wireless localization and sensing, which provide accurate position data, and high-density node deployments alongside large-scale MIMO configurations, which offer granular channel information. 
%Such wealth of information has been effectively applied in various applications, including fingerprint localization~\cite{levie2021radiounet}, power control~\cite{van2020power}, and hybrid beamforming~\cite{wu2023environment}. 
Notably, many studies have advanced the concept of knowledge maps for radio channel environments, akin to digital twins, to construct datasets representing diverse channel attributes, such as radio map~\cite{levie2021radiounet} and channel knowledge map~\cite{wu2023environment}. These developments present new opportunities to integrate environmental channel information into SIP systems for enhanced performance.

Building on these insights, this paper proposes a novel channel-aware SIP (CaSIP) framework to jointly optimize PDP allocation and PDI elimination for enhanced data detection performance in multiuser MIMO systems. Specifically, our framework optimizes user-specific, RE-wise PDP factors and devises an efficient SIP receiver network by leveraging path gain, a form of large-scale CSI. To properly leverage the path gain, we develop a path gain embedding generator in a U-Net-based channel estimation module, with the resulting embeddings incorporated into intermediate channel estimate features. Simulations validate the CaSIP framework outperforms traditional pilot (TP) schemes and state-of-the-art SIP schemes, particularly in challenging high-mobility and low-SNR scenarios.

\textit{Notations}: Boldface symbols denote matrices or vectors. $\circ$ denotes the Hadamard product and diag$(\cdot)$ transforms a vector to a diagonal matrix. $\bf 1$ is a vector of all ones and $\mathbf{I}_N$ is the $N\times N$ identity matrix. $[\cdot]_\text{indices}$ refers to a slice of a matrix or tensor, specified by a tuple of indices.
%与此同时，环境感知通信is a promising paradigm in next-generation wireless networks，其中通信环境中的信道信息，比如收发端位置、传播路径信息，被用来促进CSI的获取~\cite{zeng2024tutorial}。这对SIP方案，它在获取CSI时需要处理导频数据干扰，有特殊的意义。advancements in next-generation wireless networks make it possible to acquire high-quality channel information. specifically, Techniques such as wireless localization and sensing enable precise acquisition of position information, while high-density node deployments and large-dimensional MIMO configurations provide granular channel information.  现有无线应用充分利用这些信道信息获得成功，比如user-cell site association\cite{bethanabhotla2016optimal}，fingerprint Localization~\cite{levie2021radiounet}, power control\cite{van2020power}, hybrid beamforming~\cite{wu2023environment}。其中，一些研究根据无线信道环境构建了一种地图，形成一个数据集，里面表征各种属性的信道信息。这类似数字孪生的思想。These developments present opportunities to integrate environmental channel information into SIP schemes for improved performance.	
\section{System Model and Problem Formulation}\label{sec2}

We consider a multiuser uplink system consisting of one $M$-antenna base station (BS) and $K$ single-antenna users. The system implements an orthogonal frequency division multiplexing (OFDM) transmission scheme with $S$ subcarriers and $T$ consecutive OFDM symbols, populating a time-frequency resource grid with $E=ST$ resource elements (REs).

\subsection{SIP Scheme}\label{subsec:sipscheme}
In the SIP scheme, the transmitted signal is constructed as a superimposition of pilot symbols and data symbols, weighted by a PDP factor across the OFDM resource grid. For ease of notation, these components are defined at the granularity of individual REs rather than over a time-frequency resource grid. Specifically, for the $k$th user, let $\boldsymbol{\varphi}_k\in \mathbb{C}^{E}$ denote the pilot symbol vector, $\mathbf d_k\in \mathbb{C}^{E}$ the data symbol vector, and $\boldsymbol{\rho}_k\in \mathbb{R}^{E}$ the PDP factor vector. Then, the transmitted signal $\mathbf{s}_k$ is expressed as
\begin{align}\label{s=}
		\mathbf{s}_k=\sqrt{\mathbf{p}_k}\circ {\bm \varphi}_k + \sqrt{\tilde{\mathbf{p}}_k}\circ \mathbf{d}_k.
\end{align}
Here, $\mathbf{p}_k=P{\bm \rho}_k$ and $\tilde{\mathbf{p}}_k=P(\mathbf{1}-{\bm \rho}_k)$ denote the pilot and data power of the $k$th user, respectively. $P$ denotes the user transmitting power. Note that the PDP factor of the $k$th user at the $e$th RE $\rho_{k,e}\triangleq [{\bm \rho}_k]_{e}$ lies within the range $[0,1]$. Typically, the length of ${\bm \varphi}_k$ exceeds $K$, as practical systems often consider hundreds or thousands of REs\cite{muntan2024optimal,aoudia2021end,gu2024learning}. Hence, it ensures the availability of a sufficient number of mutually orthogonal pilot sequences where this orthogonality condition satisfies ${\bm \varphi}^\text{H}_k{\bm \varphi}_k=E$ and ${\bm \varphi}^\text{H}_k{\bm \varphi}_{k'}=0$ for $k'\neq k$. To satisfy the power constraint, the symbols for each user meet $\mathbf{x}_k^{\rm H} \mathbf{x}_k/E=1$, where ${\bf x}_k \in \{{\bf s}_k, {\bm \varphi}_k, {\bf d}_k\}$.
	
In a multiuser uplink scenario, the baseband input-output relation is given as
\begin{align}\label{y=}
	{\bf Y}=\sum_{k=1}^K {\bf H}_k  \text{diag}(\mathbf{s}_k) + {\bf N},
\end{align}
where $\mathbf{Y}\in \mathbb{C}^{M\times E}$ is the received signal matrix. The channel matrix $\mathbf{H}_k\in \mathbb{C}^{M\times E}$ represents the baseband equivalent channel of $k$th user, where each row captures the frequency-domain channel response between a specific receive antenna and user $k$. The term $\mathbf{N}\in \mathbb{C}^{M\times E}$ is the additive noise with energy $\sigma^2$. The channel matrices of all users are combined into a tensor $\boldsymbol{\mathsf{H}}\in \mathbb{C}^{M\times K\times E}$. 
	
At the BS, the receiver processes the received signal for channel estimation and signal detection. Specifically, given prior pilot knowledge of $\mathbf{p}_k$ and ${\bm \varphi}_k$, we first apply a channel estimator $f_\text{c}$ to obtain channel estimate $\hat{\mathbf H}_k$ from $\bf Y$ for each user. For instance, $f_\text{c}$ is implemented using the typical least squared (LS) algorithm, yielding
\begin{align}\label{eq:hls=}
		\hat{\mathbf{H}}_k=f_\text{c}(\mathbf{Y},\mathbf{p}_k,{\bm \varphi}_k)=\mathbf{Y}\text{diag}(\sqrt{\mathbf{p}_k}\circ {\bm \varphi}_k)^{-1},
\end{align}
where $\hat{\mathbf{H}}_k$ is the estimated channel of the $k$th user. All channel estimates are combined into a tensor $\hat{\boldsymbol{\mathsf{H}}}\in \mathbb{C}^{M\times K\times E}$. Then, the pilot interference in $\bf Y$ is subtracted, and a data detector $f_\text{d}$ is employed to recover the data symbol $\hat{\mathbf d}_k$ using $\hat{\boldsymbol{\mathsf{H}}}$. The received signal after the pilot interference elimination is expressed as  
\begin{align}\label{eq:y_s/p=}
	\mathbf{Y}_\text{s/p}={\bf Y}-\sum_{k=1}^K \hat{\bf H}_k\text{diag}(\sqrt{\mathbf{p}_k}\circ {\bm \varphi}_k).	
\end{align}   
When using the minimum mean-squared error (MMSE) algorithm for $f_\text{d}$, we recover the data symbols, computed as
\begin{align}\label{eq:hatd=}
	\mathbf{d}_{e} = f_\text{d}(\mathbf{Y}_\text{s/p},\tilde{{\bf p}}_k,\boldsymbol{\hat{\mathsf{H}}}) = \frac{\hat{\mathbf{H}}_\text{d}^\mathrm{H}}{\hat{\mathbf{H}}_\text{d}^\mathrm{H}\hat{\mathbf{H}}_\text{d}+\sigma^2\mathbf{I}}\mathbf{y}_\text{d},
\end{align}
where $\mathbf{d}_{e} \in \mathbb{C}^{K}$ is the detected data symbol vector of all user at the $e$th RE, $\hat{\bf H}_\text{d}\triangleq\sqrt{\tilde{{\bf p}}_k}[\hat{\boldsymbol{\mathsf{H}}}]_{:,:,e} \in \mathbb{C}^{M\times K}$ represents the channel between the BS and $K$ users at the $e$th RE, and $\mathbf{y}_\text{d}\triangleq[\mathbf{Y}_\text{s/p}]_{:,e} \in \mathbb{C}^M$ denotes the received signal at the $e$th RE. The detected data symbols for all users across all REs are denoted by $\hat{\mathbf{D}}\triangleq[\mathbf{d}_{1},\ldots,\mathbf{d}_{E}]^\mathrm{T} \in \mathbb{C}^{E\times K}$, with each column, denoted by $\hat{\bf d}_k$, corresponding to user $k$. 

Considering the significant errors in SIP channel estimation caused by multiuser data interference, which degrade the performance of both channel estimation and subsequent data detection, we additionally introduce the popular ICEDD algorithms. The resultant modification primarily lies in its channel estimation process, where the received signal undergoes data interference elimination. 
%in \eqref{eq:hls=} is updated as 
%	\begin{align}\label{eq:y_s/d=}
%		\mathbf{Y}^{(i)}_\text{s/d}=\mathbf{Y}-\sum_{k=1}^K \hat{\bf H}^{(i-1)}_k\text{diag}\Big(\sqrt{\tilde{{\bf p}}_k} \circ \hat{\bf d}^{(i-1)}_k\Big),
%	\end{align}
%	where the superscript $(i)$ represents the $i$th iteration.
%ICEDD~\cite{jing2018superimposed,qian2023enhancing,gan2024bayesian}

\subsection{Problem Formulation}
The objective of this paper is to enhance data detection performance in the multiuser MIMO SIP system. To this end, several issues need to be carefully examined.
First, determining the pilot and data power involves a fundamental trade-off\cite{qian2023enhancing}. Unlike traditional pilot schemes where pilot symbols occupy individual resources and are allocated full power for superior channel estimation, the SIP scheme requires a careful balance pilot and data power. This is because the superimposed data symbols must maintain adequate power for effective data detection. Consequently, identifying the optimal PDP factor emerges as a critical issue, especially in mobile multipath channels. 
%The second issue relates to inter-user interference. The superposition of pilot and data symbols amplifies interference between users, making it essential to investigate the optimal transmitted power for each user to mitigate this effect.
Second, the nonlinear PDI in the SIP scheme undermines the performance of traditional linear algorithms, such as LS and MMSE. Advanced methods are required to address this nonlinearity.
	
Building on these observations, we consider a data detection performance optimization problem that jointly optimizes the PDP factor and strategies for eliminating the nonlinear PDI. The optimization problem is formulated as
\begin{equation}\label{eq:obj=}
	\begin{split}
		\min_{\{{\bm \rho}_k, \forall k\in \mathcal{K}\},f_\text{c},f_\text{d}} \ \ &\sum_k \lVert \mathbf{d}_k-\hat{\mathbf{d}}_k\rVert^2_2\\
		\mathrm{s.t.} \quad \qquad    &0\leq [\boldsymbol{\rho}_{k}]_{e} \leq 1,\ \forall k\in \mathcal{K}, e\in \mathcal{E},
	\end{split}
\end{equation} 
where $\mathcal{K}=\{1,\ldots,K\}$ and $\mathcal{E}=\{1,\ldots,E\}$ denote the sets of users and REs, respectively. The objective function minimizes the mean squared error between the transmitted data symbols $\mathbf{d}_k$ and their detected counterparts $\hat{\mathbf{d}}_k$ across all users.

It is challenging to solve the problem \eqref{eq:obj=} due to the high dimensionality of the optimization variable ${\bm \rho}_k$ and the involvement of functional optimization~\cite{liu2020optimizing}. To address these challenges, we develop a channel-aware learning framework that fully leverages channel information to achieve a practical solution. Specifically, we introduce a power control function $f_\text{p}$, which determines the PDP factor. Additionally, the channel information is incorporated into the SIP receiver for enhanced data detection performance. The learnable parameters of $f_\text{p}$, $f_\text{c}$, $f_\text{d}$ are denoted as $f_\text{p}(\cdot;\mathbf W_\text{p})$, $f_\text{c}(\cdot;\mathbf W_\text{c})$, and $f_\text{d}(\cdot;\mathbf W_\text{d})$, respectively.

% \{{\bm \rho}_k,\forall k\},\{,\forall k\},	
% f_\text{p}(\cdot;\mathbf W_\text{p}),f_\text{c}(\cdot;\mathbf W_\text{c}),f_\text{d}(\cdot;\mathbf W_\text{d})
	
% 为了高效地得到和可行的solution，我们提出一个端到端学习架构，其中充分利用了系统的信道信息。具体地，我们引入一个功率控制函数$f_\text{p}$, 这是一个信道信息的函数，决定了导频数据功率占比和发射功率。此外，我们利用信道信息来辅助SIP接收机设计。注意$f_\text{p}$，$f_\text{c}$，$f_\text{d}$的可学习参数分别定义为,,。	  

\section{Joint Power Allocation and Receiver Optimization Leveraging Channel Information}\label{sec3}
	In this section, we present a channel-aware learning framework that integrates path gain as channel information into the end-to-end optimization process, describe the design of the CaSIP network, and explain the end-to-end training strategy to achieve the optimal data detection performance.
\begin{figure*}
	\centering
	\includegraphics[width=0.8\textwidth]{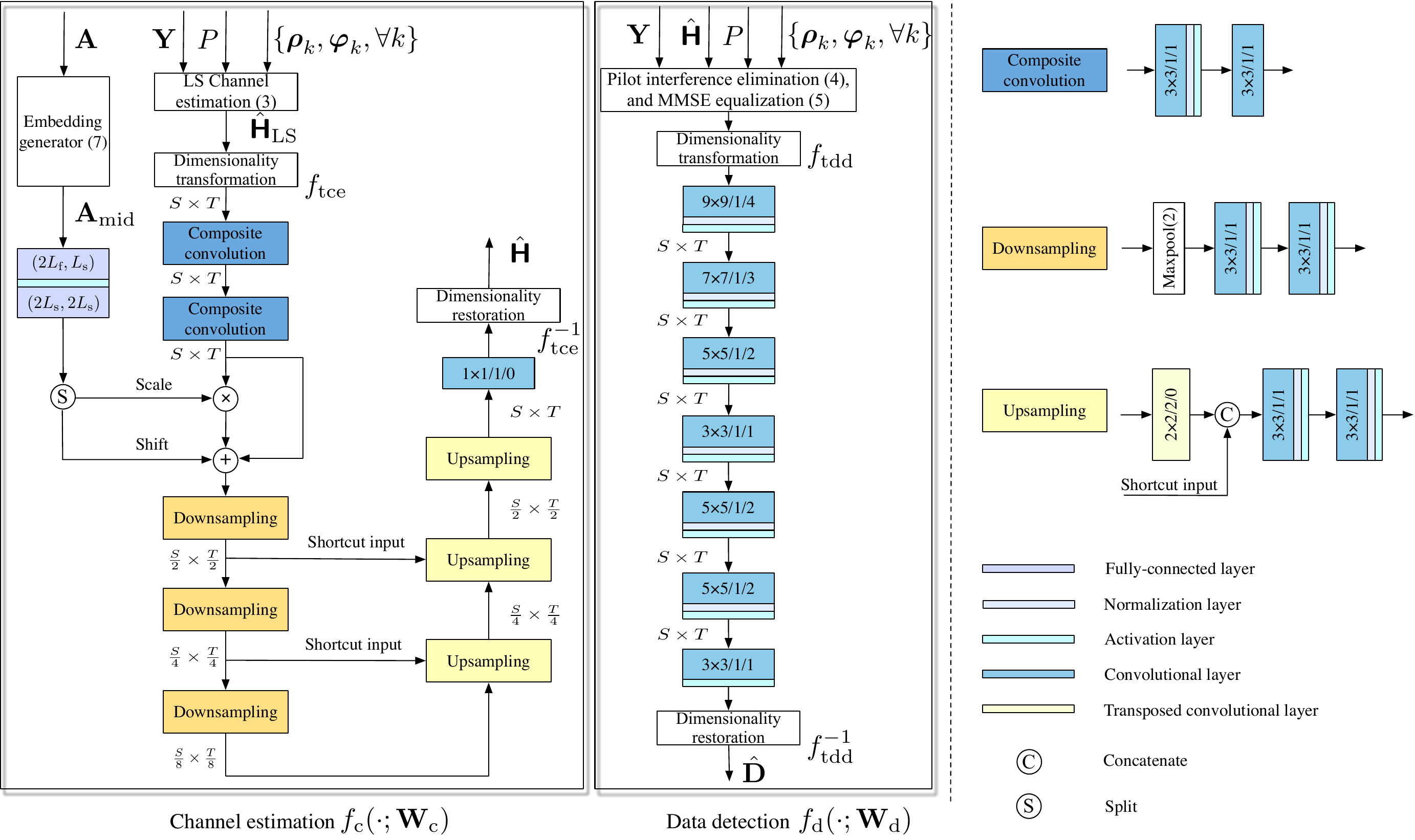}
	\caption{The receiver network of CaSIP. The parameters of fully-connected (FC) layers are denoted in the format of (input size, output size). The parameters of (transposed) convolutional layers are represented in the format of kernel size/stride/padding. 
In the channel estimation network $f_\text{c}$, SiLU is used in the MLP, while other layers use Leaky ReLU with a slope of $0.3$. In the data detection network $f_\text{d}$, ReLU is applied except for the final layer, which uses Tanh.}
	\label{fig:net}
	\vspace{-0.4cm}
\end{figure*}

\subsection{Channel-aware Learning Framework} 
Granular communication environmental information has become increasingly attainable due to advancements in wireless technologies and high-density, high-dimensional radio sampling points. In this paper, we consider the path gain of wireless links as the primary channel information, which measures the difference of signal strength between a transmitter and receiver due to large-scale effects\cite{levie2021radiounet}. 

This choice is motivated by two key factors. First, as a form of large-scale CSI, the path gain is primarily influenced by the propagation environment, making it relatively stable against the environment dynamics and predictable with minimal overhead~\cite{xu2024much}. Second, in multiuser systems, users exhibit varying path gain values due to differences in their distances from the BS. Recall that each user’s channel estimate includes the true channel response, interference from pilot signals of other users, interference from all users’ data symbols, and noise terms~\cite{qian2023enhancing}. The path gain information is critical in such scenarios as it provides a robust reference for delineating the strength of the true channel response amidst these interferences. 

Integrating the path gain information into the end-to-end optimization forms the foundation for the channel-aware learning framework, enabling effective PDP allocation and efficient PDI elimination. The specific network implementation methods are detailed in the following subsection. 

%\begin{table}[t]
%\centering
%\caption{NN Structure for Channel Estimation Refinement}
%\label{tab:cenet}
%%\resizebox{\linewidth}{!}{%
%\renewcommand\arraystretch{1.3}
%\setlength{\tabcolsep}{6mm}{
%\begin{tabular}{crc}
%%\toprule
%\hline\noalign{\hrule height 0.5pt}
%\multicolumn{3}{l}{\textbf{Input}:~$\boldsymbol{\mathsf{H}}_\text{in} \in \mathbb{R}^{N_\text{bz}KM \times 2 \times N_\text{s}\times N_\text{t}} $} \\ \hline
%\multicolumn{3}{l}{\textbf{Convolutional layers}} \\
%\textbf{Layer} & \textbf{Filters/Stride/Padding} & \textbf{Activation} \\
%(1) &$32\times 9\times 9/1/4\quad $  & BN+Tanh \\
%(2) &$64\times 5\times 5/1/2\quad $  & BN+Tanh \\
%(3) &$128\times 3\times 3/1/1\quad $  & BN+Tanh \\
%(4) &$128\times 3\times 3/1/1\quad $  & BN+Tanh \\
%(5) &$64\times 3\times 3/1/1\quad$  & BN+Tanh \\
%(6) &$32\times 5\times 5/1/2\quad$  & BN+Tanh \\
%(7) &$2\times 9\times 9/1/4\quad$  & BN \\ \hline
%\multicolumn{3}{l}{\textbf{Output}:~$\hat{\boldsymbol{\mathsf{H}}}\in \mathbb{C}^{N_\text{bz}\times K\times M\times N_\text{RE}}$}\\
%%\bottomrule
%\hline\noalign{\hrule height 0.5pt}
%\end{tabular}% 
%}
%\vspace{-0.5cm}
%\end{table}
% \mathbf{Y}\in \mathbb{C}^{M\times N_\text{RE}}, \mathbf{A}\in \mathbb{R}^{K\times N_\text{RE}}, \text{and}\ \mathbf{P}\in \mathbb{R}^{K\times N_\text{RE}} \mathbf{H}_\text{LS}\in \mathbb{C}^{N_\text{bz}\times K \times M\times N_\text{RE}} \rightarrow 
\vspace{-10pt}
\subsection{Design of CaSIP Framework}
The proposed CaSIP network leverages domain knowledge of superimposition, channel estimation, and data detection as described in Section~\ref{subsec:sipscheme}, which consists of the power control module $f_\text{p}$, the channel estimator $f_\text{c}$, and the data detector $f_\text{e}$.
\subsubsection{Power Control Module}
 At the transmitter, $f_\text{p}$ is implemented as a sigmoid function to compute the PDP factors using the input learnable parameters $\mathbf{W}_\text{p}$. This ensures that the allocated power remains within the constrained range while balancing channel estimation and data detection performance. The learned PDP factors are used in both the symbol superimposition and signal reception processes.
 
\subsubsection{Channel Estimator Model}
At the receiver, $f_\text{c}$ is designed to integrate the path gain information for channel estimation. As shown in the left part of Fig.~\ref{fig:net}, the $f_\text{c}$ backbone network is a U-Net architecture, in which down-sampling and up-sampling blocks enable multi-scale feature extraction, and shortcut connections ensure feature domain preservation~\cite{xie2023disentangled}.

We first compute the LS channel estimates as an initial approximation, denoted by $\hat{\boldsymbol{ \mathsf{H}}}_\text{LS}\in \mathbb{C}^{M\times K\times E}$. A dimensionality transformation is applied to $\hat{\boldsymbol{ \mathsf{H}}}_\text{LS}$ that stacks the real and imaginary parts and flattens the user and antenna dimensions into a new dimension, which serves as the batch dimension during training. This transformation is represented by the function $f_\text{tce}:\mathbb{C}^{M\times K\times E}\rightarrow \mathbb{R}^{MK\times 2\times S \times T}$. This step facilitates efficient 2D real-valued convolution operations and enhances the network generalizability by focusing on time-frequency domain correlations rather than user-antenna spatial relationships. 

To effectively utilize path gain, we propose a fusion mechanism that aligns the initial channel estimate features with path gain embeddings. On one hand, composite convolutional blocks are applied to the reshaped LS estimates before down-sampling. These front convolutional blocks transform the initial channel estimates into a more informative feature space by increasing feature channels, preparing them for subsequent refinement stages. This early fusion in the pipeline also ensures that both amplitude and structural relationships are preserved and propagated through the down-sampling and up-sampling stages. On the other hand, we design a path gain embedding generator that produces positional embeddings with both scale and shift components. Specifically, the embeddings are inspired by the sinusoidal positional embedding method, which is particularly effective for capturing relative positional relationships~\cite{vaswani2017attention}. An additional multi-layer perceptron (MLP) extracts these scale and shift components.

 Given $N_\text{bz}$ path gain values, each represented by $\mathbf{A}\in \mathbb{R}^{M\times K}$, we first compute the immediate features that incorporate positional information using its reshaped form $\mathbf{a}\in \mathbb{R}^{M K}$, which is expressed by
 \begin{align}
 	\mathbf{A}_\text{mid}=\mathrm{Concatenate}\big(\sin(\boldsymbol{\mathbf{a}}\mathbf{f}_\text{s}^\mathrm{T}),\cos(\boldsymbol{\mathbf{a}}\mathbf{f}_\text{s}^\mathrm{T})\big),
 \end{align}
where $\mathbf{f}_\text{s}$ is a frequency scaling vector with a length of $L_\text{f}$, defined by $[\mathbf{f}_\text{s}]_i=\exp^{-\log(10000)(i-1)/L_\text{f}}, i=1,\ldots,L_\text{f}$. By concatenating the sine and cosine components along the feature dimension, we obtain $\mathbf{A}_\text{mid} \in \mathbb{R}^{M K\times 2L_\text{f}}$. Then we generate the path gain embeddings using an MLP, which is further split equally into scale and shift components along the feature dimension, with each component having a length of $L_\text{s}$. These scale and shift values are subsequently applied to adjust the channel estimates, allowing the network to dynamically refine predictions based on the path gain information.

The fused features, enriched by path gain embeddings, are processed through down-sampling and up-sampling blocks. After the dimensionality restoration, inverse version of $f_\text{tce}$ and denoted by $f^{-1}_\text{tce}$, the final refined channel estimates  $\hat{\boldsymbol{\mathsf{H}}}$ are output by the network.

\subsubsection{Data Detector Model}
Following channel estimation, we design the $f_\text{d}$ network to obtain data estimates by removing residual noise and PDI. As illustrated in the middle part of Fig.~\ref{fig:net}, this module employs multiple convolutional layers, normalization layers, and activation layers to refines the MMSE equalized data. A dimensionality transformation similar to $f_\text{tce}$ is applied to the MMSE equalized data, denoted by the function $f_\text{tdd}:\mathbb{C}^{K\times E}\rightarrow \mathbb{R}^{K\times 2\times S \times T}$. The width and height dimensions of the features are kept identical to those of the network input in each convolutional layer, preserving the spatial structure of the reshaped equalized input. The final data estimate $\hat{\bf D}$ is processed using the inverse transformation of $f_\text{tdd}$, denoted by $f^{-1}_\text{tdd}$.

\vspace{-15pt}
\subsection{Training Details}
The CaSIP network is trained in an end-to-end manner to minimize the MSE between the transmitted symbols and the estimated symbols. All learnable parameters are optimized by
\begin{align}\label{eq:loss=}
		(\mathbf{W}_\text{p}^\ast,\mathbf{W}_\text{c}^\ast,\mathbf{W}_\text{e}^\ast)= \argmin_{\mathbf{W}_\text{p},\mathbf{W}_\text{c},\mathbf{W}_\text{e}} \sum_k \lVert \mathbf{d}_k-\hat{\mathbf{d}}_k\rVert^2_2.
	\end{align} 

During forward propagation, data symbols $\mathbf{d}_k$ with a specific modulation scheme are randomly generated, and pilot symbols $\boldsymbol{\varphi}_k$ are selected from a discrete Fourier transform (DFT) matrix. The $e$th element of the $k$th user’s pilot symbols ${\bm \varphi}_k$ is specified as $\exp^{-j2\pi(k-1)(e-1)/K}$ and ${\bm \varphi}_k$ is placed in Fortran-like order on the time-frequency resource grid. Gradients are back-propagated to optimize all network parameters. As training converges, the CaSIP framework can learn the optimal PDP factors and SIP neural receiver, achieving the enhanced data detection performance.
\section{Simulation Results}
\begin{table}[t]
		\centering
		\caption{Parameters Used for Training and Evaluation }
		\renewcommand\arraystretch{1.1}
		\begin{tabular}{l|c|c}
			\hline\noalign{\hrule height 0.5pt}
			Parameter  & Symbol &Value \\ \hline
			Transmit power & $P$ & $20$ dBm\\ \hline
			Number of UEs   &$K$ & $12$\\\hline
			Number of BS antennas   &$M$ & $64$\\\hline
			Number of OFDM symbols  &$N_\text{t}$ & $14$ ($1$ slot) \\\hline
			Number of subcarriers   &$N_\text{s}$ & $48$ ($4$ RBs) \\\hline
			Number of REs & $E$ & $672$ \\ \hline
			Code rate  &/& $490/1024$ (LDPC)\\ \hline
			Modulation &/& $16$-QAM\\\hline
			Delay profile & /  & CDL-A \\ \hline
			Carrier frequency  & /  & $2.6$~GHz\\ \hline
			Subcarrier spacing  & / & $30$~kHz\\\hline
			Mobility velocity & $v$  & $3/60$~km/h \\\hline
			Delay spread range for training & / & $100$ to $300$~ns\\ \hline
			Delay spread used for test & / & $200$~ns \\ \hline
			$E_s/\sigma^2$ range for training & / & $10$ to $14$~dB\\ \hline\noalign{\hrule height 0.5pt}
		\end{tabular}
		\label{tab:simu_paras}
%		\vspace{-0.5cm}
	\end{table}
	\subsection{Simulation Settings}
We construct a realistic channel dataset using the 3GPP CDL-A delay profile, considering two mobile velocity scenarios and a dynamically varying delay spread within a specific range. A total of $900$ channel samples are generated, with user locations categorized into $45$ classes, randomly distributed between $150$~m and $600$~m from a BS positioned at a height of $30$~m. The dataset is split into training, validation, and testing sets in a $7:1:1$ ratio. This paper assumes that channel gains between users and receive antennas, which can be derived from the channel samples, are available as network inputs. The parameters $L_\text{f}$ and $L_\text{s}$ are set to $8$ and $32$, respectively. Network training employs the Adam optimizer with a learning rate of $10^{-4}$~($\beta_1=0.9$ and $\beta_2=0.99$) over $300$ epochs. The CaSIP network comprises $2.65$ million parameters.

Our evaluation emphasizes system sum throughput as the primary metric for assessing data detection performance. This choice aligns with the practical requirements of digital communication systems and is further justified by the significant gains observed in ICEDD algorithms that integrate channel decoding~\cite{qian2023enhancing}. Accordingly, we test the proposed scheme in a link-level system comprising an Uplink Shared Channel simulator and a Physical Uplink Shared Channel simulator. The simulation parameters are outlined in Table~\ref{tab:simu_paras}. 
	For a given set of OFDM and channel parameters, the average energy of the received symbols to noise power ratio is defined as $E_s/\sigma^2= P\mathbb{E}\big\{||\sum_k\mathbf{H}_k||^2\big\}/ \sigma^2$.
	% ~\cite{3gpp38212,3gpp38211}

	\subsection{Performance Evaluation}
	We evaluate the proposed approach against three categories of baseline schemes. The first baseline, TP, follows the 5G NR DM-RS Type A configuration Type II, which reserves two OFDM symbols for pilots to perform standard channel estimation~\cite{3gpp38211}. The second category includes the SIP scheme and its variant that assumes perfect CSI acquisition. These schemes uniformly adopt a PDP factor of $0.3$ across all users and REs while incorporating the ICEDD algorithms, where LDPC-decoded bit information is used to refine channel estimation~\cite{qian2023enhancing}. Finally, the SIPCE scheme jointly optimizes the RE-wise PDP factors and channel estimation with a weight of $10$ for the latter, followed by implementing the ICEDD framework for further PDI interference~\cite{gu2024learning}. Note that the SIPCE scheme shares the proposed superimposition and channel estimation networks for fair comparison, where its parameter count is reduced by $0.72$ million due to the absence of a data detection network.
	
	\begin{figure}[t]
		\centering
		\hspace{-0.3cm}
		\subfigure[$v=3$~km/h]{
			\label{fig:thp_v3}
			\includegraphics[width=0.48\linewidth]{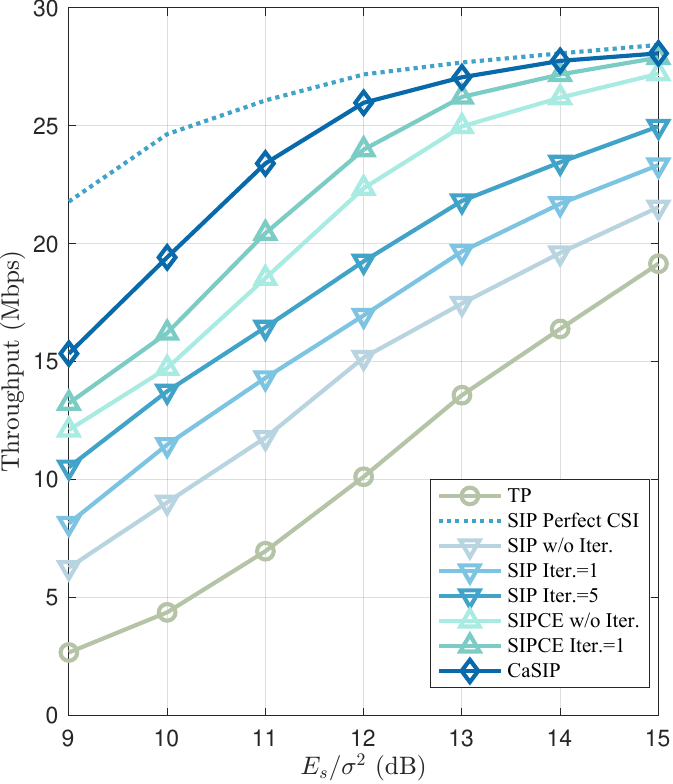} % sumthp_v3_Nov.pdf
		}
		\hspace{-0.3cm}
		\subfigure[$v=60$~km/h]{
			\label{fig:thp_v60}
		\includegraphics[width=0.48\linewidth]{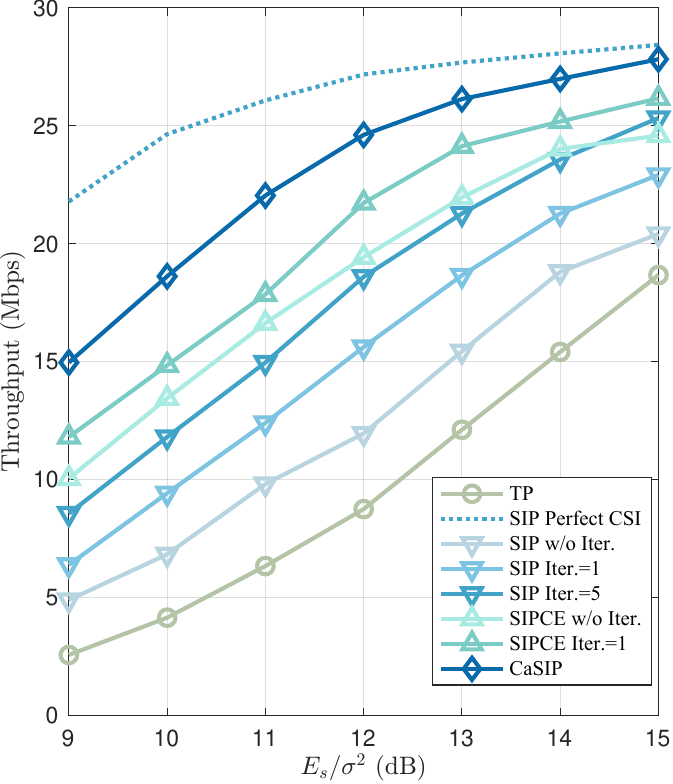}
		}
		\caption{Sum throughput versus $E_s/\sigma^2$.}
		\label{fig:thp}
		\vspace{-0.3cm}
	\end{figure}
	
		\begin{figure}[t]
		\centering
		\hspace{-0.3cm}
		\subfigure[$v=3$~km/h]{
			\label{fig:nmse_v3}
			\includegraphics[width=0.48\linewidth]{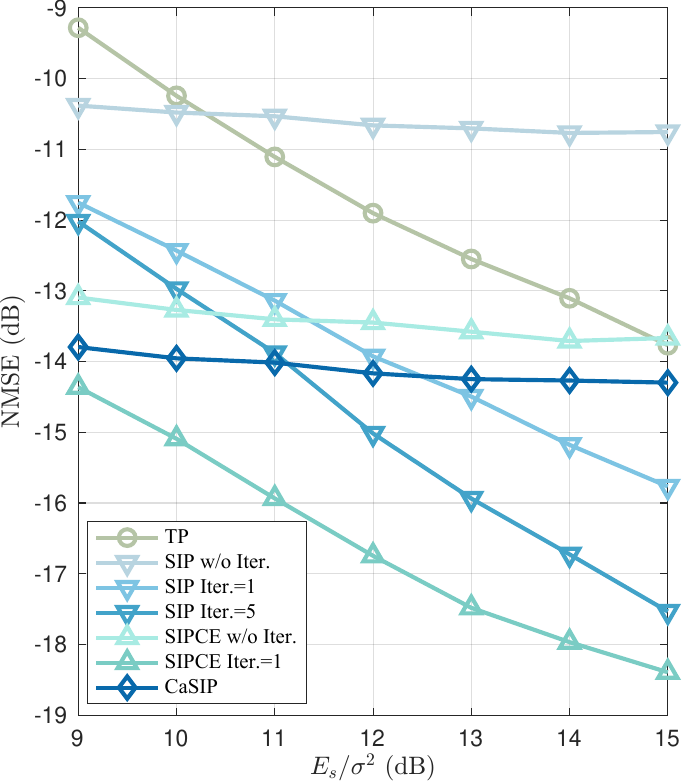}
		}
		\hspace{-0.3cm}
		\subfigure[$v=60$~km/h]{
			\label{fig:nmse_v60}
			\includegraphics[width=0.48\linewidth]{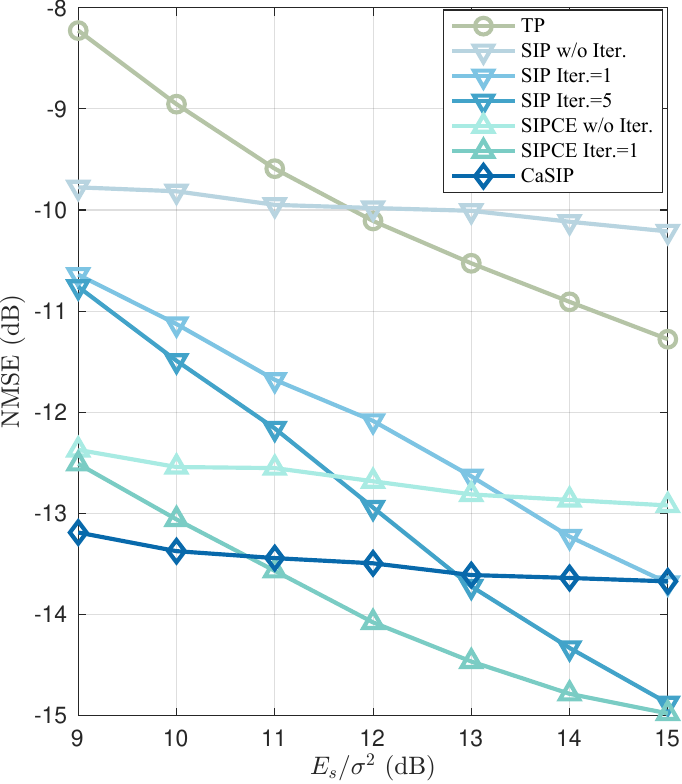}
		}
		\caption{Channel estimation NMSE versus $E_s/\sigma^2$.}
		\label{fig:nmse}
		\vspace{-0.3cm}
	\end{figure}

	\begin{figure}[t]
		\centering
		\includegraphics[width=\linewidth]{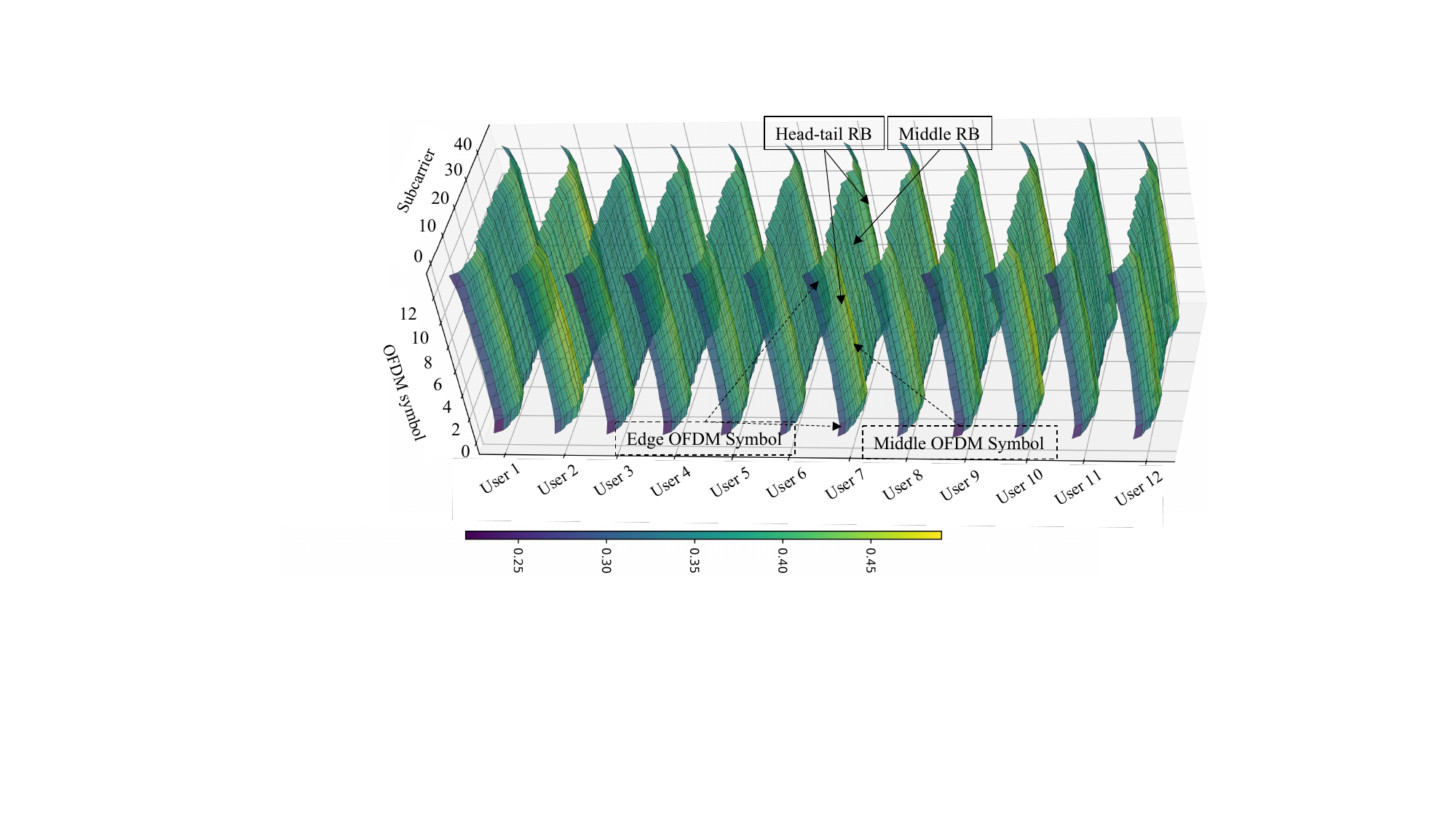}
		\caption{Heat maps of the learned PDP factors for all users at $v=3$~km/h.}	
		\label{fig:pwr_factor}		
		\vspace{-0.3cm}
	\end{figure}
	
\begin{table}[t]
	\centering
	\caption{The Learned PDP Factor (\%) Variations of CaSIP by Mobility levels.}
	\label{tab:pwr_factor}
	\renewcommand{\arraystretch}{1.5}
	\resizebox{\columnwidth}{!}{%
	\begin{tabular}{c|c|c|c|c|c}
	\hline
	Mobility & Whole RBs & Head-tail RBs & Middle RBs & Edge Symbols & Middle Symbols \\ \hline
	$3$~km/h & $38.97_{\pm 3.55}$ & $39.22_{\pm 4.55}$ & $38.71_{\pm 2.10}$ & $37.23_{\pm 3.81}$ & $39.66_{\pm 3.19}$ \\ \hline
	$60$~km/h & $39.67_{\pm 3.72}$ & $40.65_{\pm 4.69}$ & $38.68_{\pm 1.92}$ & $37.95_{\pm 3.91}$ & $40.35_{\pm 3.40}$ \\ \hline
%	SIPCE at $60$~km/h & $35.97_{\pm 1.21}$ & $36.00_{\pm 1.28}$ & $35.95_{\pm 1.12}$ & $35.81_{\pm 1.19}$ & $36.03_{\pm 1.21}$ \\ \hline
%	SIPCE at $3$~km/h & $36.41_{\pm 1.24}$ & $36.41_{\pm 1.32}$ & $36.41_{\pm 1.15}$ & $36.26_{\pm 1.21}$ & $36.48_{\pm 1.24}$ \\ \hline
	\end{tabular}%
	}
%	Mobility & TB & head-tail RB & middle RB & edge OFDM symbol & middle OFDM symbol \\ \hline
%	$3$ & $38.97_{\pm 0.04}$ & 0.3922, std = 0.05 & 0.3871, std = 0.02 & 0.3723, std = 0.04 & 0.3966, std = 0.03 \\ \hline
%	$60$ & 0.3967, std = 0.04 & 0.4065, std = 0.05 & 0.3868, std = 0.02 & 0.3795, std = 0.04 & 0.4035, std = 0.03
\end{table}
% & Whole RB & Head-tail RB & Middle RB & Edge OFDM symbol & Middle OFDM symbol
	 
	Fig.~\ref{fig:thp_v3} compares the sum throughput performance of all schemes under low-mobility scenarios. CaSIP achieves the highest performance efficiently, requiring no iterative processing, and even approaches the performance of perfect CSI under high-SNR conditions. We further present in Fig.~\ref{fig:nmse_v3} the normalized mean squared error (NMSE) performance of channel estimation under low-mobility scenarios. Although both SIPCE and SIP leverage the ICDEE algorithms to eliminate PDI and refine channel estimation, as seen in Fig.~\ref{fig:nmse_v3}, CaSIP still outperforms their iterative outputs. This result highlights the effectiveness of the channel-aware end-to-end optimization approach, which specifically targets enhanced data detection. The throughput performance improvements observed in both CaSIP and SIPCE over the baseline SIP scheme underscore the importance of jointly learning high-dimensional PDP factors and a PDI elimination-based receiver network. In contrast, TP exhibits the worst performance, primarily due to reduced data transmission resources and poor channel estimation accuracy, particularly under low-SNR conditions. 
	
	From Fig.~\ref{fig:thp_v60} and Fig.~\ref{fig:nmse_v60}, which illustrate high-mobility scenarios, it is observed that both throughput and channel estimation performance decline across all schemes due to rapid channel variations, which pose significant challenges for accurate channel estimation. Nevertheless, CaSIP experiences the least degradation in throughput performance, showcasing its robustness to dynamic channel conditions.

Fig.~\ref{fig:pwr_factor} presents the learned PDP factors in low-mobility scenarios, visualized as a heat map. Higher PDP factor values are concentrated in the head-tail resource blocks (RBs), as well as in the middle OFDM symbol positions. Specifically, these positions correspond to the first $12$ and last $12$ subcarriers within the $3$rd to $12$th OFDM symbols. This characteristic remains consistent in high-mobility scenarios. Table~\ref{tab:pwr_factor} provides the percentage-based PDP factor variations. As the velocity increases, the PDP factors in these positions increase significantly. This is attributed to the need for higher pilot power allocation to handle the challenges posed by time-varying channels.
\section{Conclusion}
This paper proposes a channel-aware DL-based SIP framework to enhance data detection performance of multiuser MIMO systems. By effectively incorporating path gain information, the proposed CaSIP enables the optimal assignment of PDP factors and develops the efficient SIP receiver network for PDI elimination. Simulation results demonstrate the effectiveness of the proposed framework under various conditions.

% or


%\appendices
%
%\input{sections/07_appendix}
%% you can choose not to have a title for an appendix
%% if you want by leaving the argument blank
%\section{}

%\section*{Acknowledgment}


\ifCLASSOPTIONcaptionsoff
  \newpage
\fi



%\bibliographystyle{./IEEEtran.bst}
%\bibliography{IEEEabrv, \rootpath/reference.bib}
%\bibliography{reference.bib}
%\bibliography{IEEEabrv, reference.bib}
%\bibliography{jabref.bib}
\bibliographystyle{IEEEtran}
\bibliography{myrefers.bib}

\begin{thebibliography}{10}
\providecommand{\url}[1]{#1}
\csname url@samestyle\endcsname
\providecommand{\newblock}{\relax}
\providecommand{\bibinfo}[2]{#2}
\providecommand{\BIBentrySTDinterwordspacing}{\spaceskip=0pt\relax}
\providecommand{\BIBentryALTinterwordstretchfactor}{4}
\providecommand{\BIBentryALTinterwordspacing}{\spaceskip=\fontdimen2\font plus
\BIBentryALTinterwordstretchfactor\fontdimen3\font minus \fontdimen4\font\relax}
\providecommand{\BIBforeignlanguage}[2]{{%
\expandafter\ifx\csname l@#1\endcsname\relax
\typeout{** WARNING: IEEEtran.bst: No hyphenation pattern has been}%
\typeout{** loaded for the language `#1'. Using the pattern for}%
\typeout{** the default language instead.}%
\else
\language=\csname l@#1\endcsname
\fi
#2}}
\providecommand{\BIBdecl}{\relax}
\BIBdecl

\bibitem{xw2023toward}
W.~Xu, Y.~Huang, W.~Wang, F.~Zhu, and X.~Ji, ``Toward ubiquitous and intelligent {{6G}} networks: {From} architecture to technology,'' \emph{Sci. China Inf. Sci.}, vol.~66, no.~3, pp. 130\,300:1--2, Mar. 2023.

\bibitem{haritha2024superimposed}
H.~Haritha, D.~N. Amudala, R.~Budhiraja, and A.~K. Chaturvedi, ``Superimposed versus regular pilots for hardware impaired rician-faded cell-free massive {MIMO} systems,'' \emph{IEEE Trans. Commun.}, vol.~72, no.~11, pp. 6688--6706, Apr. 2024.

\bibitem{zhou2024optimized}
X.~Zhou \emph{et~al.}, ``Optimized payload length and power allocation for generalized superimposed pilot in {URLLC} transmissions,'' \emph{IEEE Trans. Commun.}, vol.~72, no.~10, pp. 6073--6086, Apr. 2024.

\bibitem{xie2024superimposed}
M.~Xie \emph{et~al.}, ``Superimposed pilots for cell-free massive {MIMO} over spatial-correlated rician fading channels,'' \emph{IEEE Trans. Wireless Commun.}, early access, Nov. 2024.

\bibitem{muntan2024optimal}
I.~P. Muntan and M.~J. F.-G. Garc, ``Optimal estimation of frequency-selective channels in ofdm-based superimposed training schemes,'' \emph{IEEE Trans. Veh. Technol.}, vol.~73, no.~11, pp. 15\,969--15\,982, Jul. 2024.

\bibitem{jing2018superimposed}
X.~Jing \emph{et~al.}, ``Superimposed pilot optimization design and channel estimation for multiuser massive {MIMO} systems,'' \emph{IEEE Trans. Veh. Technol.}, vol.~67, no.~12, pp. 11\,818--11\,832, Dec. 2018.

\bibitem{gan2024bayesian}
X.~Gan, C.~Huang, Z.~Yang, C.~Zhong, X.~Chen, Z.~Zhang, Q.~Guo, C.~Yuen, and M.~Debbah, ``Bayesian learning for double-{RIS} aided {ISAC} systems with superimposed pilots and data,'' \emph{IEEE J. Sel. Top. Signal Process.}, pp. 1--16, early access, May 2024.

\bibitem{qian2023enhancing}
C.~Qian \emph{et~al.}, ``Enhancing wideband multiuser mimo uplink using superimposed pilots: Joint receiver design,'' \emph{IEEE Wireless Commun. Lett.}, vol.~13, no.~4, pp. 1138--1142, Apr. 2024.

\bibitem{xie2023disentangled}
R.~Xie \emph{et~al.}, ``Disentangled representation learning for {RF} fingerprint extraction under unknown channel statistics,'' \emph{IEEE Trans. Commun.}, vol.~71, pp. 3946--3962, Jul. 2023.

\bibitem{zeng2023multi}
Y.~Zeng, Y.~Gong, J.~Liu, S.~Lin, Z.~Han, R.~Cao, K.~Huang, and K.~B. Letaief, ``Multi-channel attentive feature fusion for radio frequency fingerprinting,'' \emph{IEEE Trans. Wireless Commun.}, vol.~23, no.~5, pp. 4243--4254, May~2024.

\bibitem{xu2023edge}
W.~Xu, Z.~Yang, D.~W.~K. Ng, M.~Levorato, Y.~C. Eldar, and M.~Debbah, ``Edge learning for {B5G} networks with distributed signal processing: Semantic communication, edge computing, and wireless sensing,'' \emph{IEEE J. Sel. Top. Signal Process.}, vol.~17, no.~1, pp. 9--39, Jan. 2023.

\bibitem{xu2024disentangled}
W.~Xu \emph{et~al.}, ``Disentangled representation learning empowered {CSI} feedback using implicit channel reciprocity in {FDD} massive {MIMO},'' \emph{IEEE Trans. Wireless Commun.}, vol.~23, no.~10, pp. 15\,169--15\,184, Jul.~2024.

\bibitem{aoudia2021end}
F.~A. Aoudia and J.~Hoydis, ``End-to-end learning for {OFDM}: From neural receivers to pilotless communication,'' \emph{IEEE Trans. Wireless Commun.}, vol.~21, no.~2, pp. 1049--1063, Feb. 2022.

\bibitem{gu2024learning}
R.~Gu \emph{et~al.}, ``Learning power allocation and channel estimation for superimposed pilot-assisted multiuser {MIMO},'' \emph{IEEE Wireless Commun. Lett.}, vol.~13, no.~11, pp. 3025--3029, Aug. 2024.

\bibitem{zeng2024tutorial}
Y.~Zeng \emph{et~al.}, ``A tutorial on environment-aware communications via channel knowledge map for {6G},'' \emph{IEEE Commun. Surv. Tutor.}, vol.~26, no.~3, pp. 1478--1519, Feb. 2024.

\bibitem{levie2021radiounet}
R.~Levie, {\c{C}}.~Yapar, G.~Kutyniok, and G.~Caire, ``Radiounet: Fast radio map estimation with convolutional neural networks,'' \emph{IEEE Trans. Wireless Commun.}, vol.~20, no.~6, pp. 4001--4015, Feb. 2021.

\bibitem{wu2023environment}
D.~Wu, Y.~Zeng, S.~Jin, and R.~Zhang, ``Environment-aware hybrid beamforming by leveraging channel knowledge map,'' \emph{IEEE Trans. Wireless Commun.}, vol.~23, no.~5, pp. 4990--5005, Oct. 2023.

\bibitem{liu2020optimizing}
D.~Liu, C.~Sun, C.~Yang, and L.~Hanzo, ``Optimizing wireless systems using unsupervised and reinforced-unsupervised deep learning,'' \emph{IEEE Netw.}, vol.~34, no.~4, pp. 270--277, Feb. 2020.

\bibitem{xu2024much}
X.~Xu and Y.~Zeng, ``How much data is needed for channel knowledge map construction?'' \emph{IEEE Trans. Wireless Commun.}, vol.~23, no.~10, pp. 13\,011--13\,021, May 2024.

\bibitem{vaswani2017attention}
A.~Vaswani \emph{et~al.}, ``Attention is all you need,'' in \emph{Proc. Int. Conf. Adv. Neural Inf. Processing Syst.}, Long Beach, CA, USA, Dec. 2017, pp. 5998--6008.

\bibitem{3gpp38211}
3GPP, ``{NR; Physical channels and modulation},'' {3rd Generation Partnership Project (3GPP)}, Technical Specification (TS) 38.211, Jan. 2023, version 17.4.0.

\end{thebibliography}
%


\end{document}